 \documentclass[final,5p,times,twocolumn]{elsarticle}

\usepackage{graphicx}

\usepackage{epsfig}
\usepackage{epstopdf}

 \biboptions{sort&compress}

\journal{Journal of Nuclear Materials}

\begin{document}

\begin{frontmatter}



\title{Plasma cleaning of ITER first mirrors in magnetic field}


\author[unibasel_phys]{Lucas Moser\corref{cor1}}
\ead{lucas.moser@unibas.ch}

\author[unibasel_phys]{Roland Steiner}

\author[ITER]{Frank Leipold}

\author[ITER]{Roger Reichle}

\author[unibasel_phys]{Laurent Marot}

\author[unibasel_phys]{Ernst Meyer}

\cortext[cor1]{Corresponding author}

\address[unibasel_phys]{Department of Physics, University of Basel, Klingelbergstrasse 82, CH-4056, Basel, Switzerland}
\address[ITER]{ITER Organization, Route de Vinon-sur-Verdon, 13115 St Paul-lez-Durance, France}

\begin{abstract}
To avoid reflectivity losses in ITER's optical diagnostic systems, plasma sputtering of metallic First Mirrors is foreseen in order to remove deposits coming from the main wall (mainly beryllium and tungsten). Therefore plasma cleaning has to work on large mirrors (up to a size of 200$\times$300 mm) and under the influence of strong magnetic fields (several Tesla). This work presents the results of plasma cleaning of aluminium and aluminium oxide (used as beryllium proxy) deposited on molybdenum mirrors. Using radio frequency (13.56 MHz) argon plasma, the removal of a 260 nm mixed aluminium/aluminium oxide film deposited by magnetron sputtering on a mirror (98 mm diameter) was demonstrated. 50 nm of pure aluminium oxide were removed from test mirrors (25 mm diameter) in a magnetic field of 0.35 T for various angles between the field lines and the mirrors surfaces. The cleaning efficiency was evaluated by performing reflectivity measurements, Scanning Electron Microscopy and X-ray Photoelectron Spectroscopy.

\end{abstract}


\end{frontmatter}
\section{Introduction}
\label{1}
Metallic First Mirrors (FMs) will play an essential role in ITER to ensure well controlled fusion reactions and proper plasma analysis. They will be the first elements of a majority of optical diagnostics, guiding the light originating from the plasma or from probing light sources through the neutron shielding towards detectors. Due to their proximity to the plasma, FMs will experience high particle fluxes (charge-exchange neutrals and neutrons but also ultraviolet, X-ray and gamma radiations) leading to erosion and/or deposition. Especially net deposition of particles eroded from the main wall, i.e. mainly beryllium (Be) and tungsten (W), can severely degrade the FMs reflectivity and therefore endanger the reliability of optical diagnostics. Insitu plasma sputtering is currently considered as one of the most promising cleaning techniques to remove deposits from FMs \cite{ITER1}. 

Porous films containing Be were already reported to grow in JET and in PISCES-B \cite{Rubel1,Temmerman1}. In our specific setup, aluminium (Al) depositions were used to simulate this kind of films to avoid Be due to its toxicity \cite{Eren1}. Al and Be have similar chemical properties \cite{Marot1}. Molybdenum (Mo) mirrors were used as metallic mirrors as they are currently considered as one of the best candidates for FMs \cite{Lit1}.

In previous works, the successful cleaning of Mo mirrors (18 mm diameter) was achieved using a radio-frequency (RF) plasma operating at a frequency of 13.56 MHz \cite{ITER2,Moser1}. It was possible to remove pure Al / Al$_{2}$O$_{3}$, pure W / W$_{oxide}$ and mixed Al / Al$_{2}$O$_{3}$ / W deposits using argon (Ar), neon and Ar + deuterium (D$_{2}$) mixture with ion energies between 150 and 350 eV while maintaining good optical properties. 

Nevertheless, the size of FMs used in ITER (for example: 200$\times$300 mm for Edge Thomson Scattering diagnostic) and the presence of ITER's permanent magnetic field (several Tesla at the first wall) may affect the plasma cleaning process. To investigate the effects of these conditions (large mirror and magnetic field), two separate and distinct set of experiments were conducted: (i) Plasma cleaning without magnetic field on a poly-crystalline Mo mirror with a diameter of 98 mm (mirror A) deposited with Al and Al$_{2}$O$_{3}$. (ii) Plasma cleaning on mirrors consisting of a stainless steel plate of 25 mm diameter and a 300 nm coating of nano-crystalline Mo \cite{Eren2} (mirror B). They are subsequently coated with dense Al$_{2}$O$_{3}$. The cleaning is performed in the presence of a magnetic field (0.35 T) where the angle between the field lines and the mirror's surface was varied from 0$^{\circ}$ to 90$^{\circ}$.

\section{Experimental Conditions}
\label{2}

\begin{figure}[h]
\centering
\includegraphics[width=0.4\textwidth]{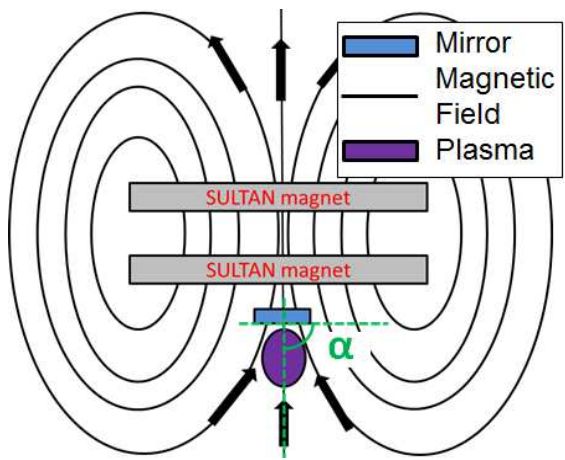}
\caption{\label{Sultan}Representation (top view) of the SULTAN facility and the position of the mirror and plasma in the magnetic field lines. The angle between the field lines and the mirrors surface is denoted $\alpha$.}
\end{figure}

Each experiment presented in this work is a two-stage operation: Firstly the deposition of the film on the mirror and secondly the removal of this film with plasma generated by applying RF directly to the mirror at a frequency of 13.56 MHz (RF capacitively coupled discharge where the mirror serves the electrode). This type of discharge leads to the formation of a negative DC component on the mirror called self-bias. This self-bias has an influence on the sputtering energy of the ions. For the first set of experiments done at the University of Basel without magnetic field (see Section \ref{WnMF}), Al films have been deposited on mirror A with magnetron sputtering as described in \cite{Eren1}. The process was done in an D$_{2}$ and Ar environment at a pressure of 3 Pa (Ar partial pressure: 18\%) where Ar was used to enhance the Al deposition rate. This leads to a similar film reported by Marot \textit{et al.} (Fig. 2 in \cite{Marot1}), relevant to what is expected in ITER. Plasma cleaning was performed with an Ar plasma (0.5 Pa) and for different ion energies (200 to 350 eV). For the second set of experiments, mirrors B were coated with pure and dense Al$_{2}$O$_{3}$. To do so, the deposition was done in an Ar and O$_{2}$ environment of 1.5 Pa (partial pressure of Ar: 50\%) with facing magnetron sputtering. Plasma cleaning in a magnetic field environment was carried out at the SULTAN facility in EPFL-CRPP Villigen (see Section \ref{WMF}) \cite{Bruzzone}. The cleaning of the Al$_{2}$O$_{3}$ film was done using Ar plasma (1.5Pa) with 200 eV ion energy (when not mentioned, 20W of RF power were needed to achieve 200 eV ion energy). The vacuum chamber was set outside of SULTAN facility where the magnetic field was 0.35 T (Fig. \ref{Sultan}). The angle, $\alpha$, between the magnetic field lines and the mirror's surface could be varied from 0$^{\circ}$ to 90$^{\circ}$. For the surface composition analysis, the mirror A was characterized by means of Energy Dispersive X-ray Photospectroscopy (EDX) with a SEM-FEI Nova Nano SEM230 at 15kV, mirrors B by means of X-ray Photoelectron Spectroscopy (XPS). The setup and fitting procedure are described elsewhere \cite{Eren3}. For both types of mirrors, total and diffuse reflectivity were measured with a Varian Cary 5 spectrophotometer (250-2500 nm) and surface morphology was investigated using a Scanning Electron Microscope (SEM) Hitachi S-4800 field emission at 5 kV. 

\section{Results and Discussion}
\label{3}
\subsection{Cleaning in the absence of a magnetic  field}
\label{WnMF}

\begin{figure}[h]
\centering
\includegraphics[width=0.4\textwidth]{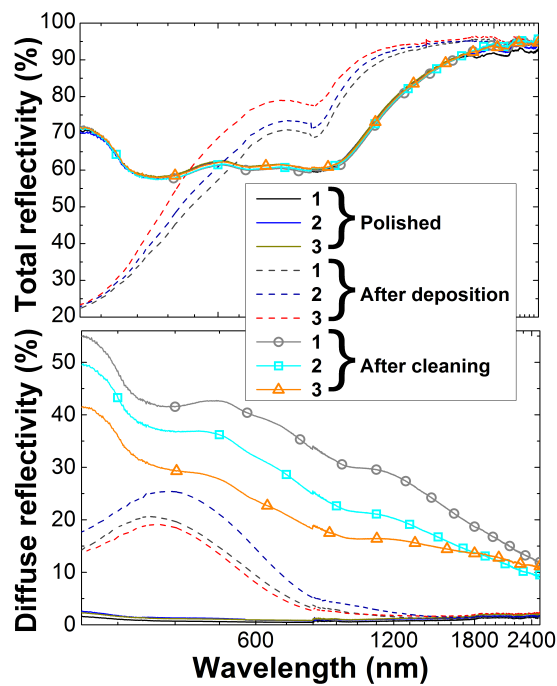}
\caption{\label{RBM}Total and diffuse reflectivity measurements of Mo polished mirror, after deposition of an Al/$Al_{oxide}$ film and after plasma cleaning. The measurements were done on position 1,2 and 3 from Fig. \ref{EDX} (a)).}
\end{figure}

\begin{figure}[h]
\centering
\includegraphics[width=0.45\textwidth]{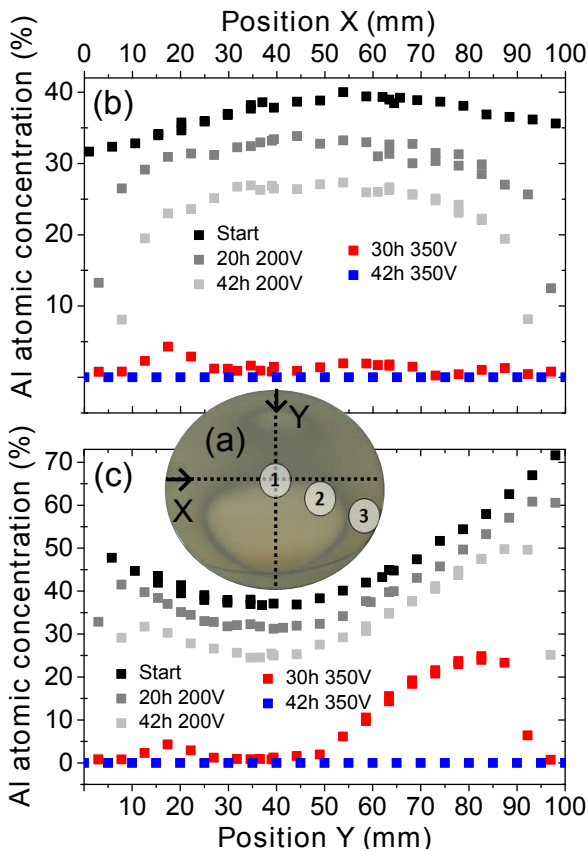}
\caption{\label{EDX}(a) Picture of the Mo mirror after the third cleaning cycle. 1,2 and 3 are the positions for the reflectivity measurements. The X and Y axis on the mirror are used for the EDX measurements. (b) and (c) EDX measurements where the Al atomic concentration are normalised by the measured Mo and Al values $\left(Al(\%) =  \frac{Al(\%)}{Al(\%)+ Mo(\%)}\right)$}
\end{figure}

The polished mirror A was deposited with a 260 nm thick Al / Al$_{2}$O$_{3}$ film. The total reflectivity of this mirror decreased drastically in the UV range compared to the polished one (Fig. \ref{RBM}). To remove the deposited film, four cleaning cycles (Ar, 0.5 Pa) were necessary. Two with a self-bias of 200 V for 20 and 42 hours, and two with a self-bias of 350 V for 30 and 42 hours. The EDX measurements carried out after each cleaning cycle (Fig. \ref{EDX}) clearly showed a homogeneous cleaning over the whole surface, i.e. the fraction of Al decreased with the same speed along the X and Y axis, except at the edge of the mirror. No electrical shielding of the mirror i.e. no metallic surrounding at ground potential was used for the cleaning: more ions were collected at the edge thus increasing the cleaning rates. The last EDX measurement showed a total removal of all Al from the mirror surface. This can also be seen by the recovering of the total reflectivity (Fig. \ref{RBM}). On the other hand, the diffuse reflectivity increased from a few percent up to 55\% after 130 hours of cleaning. Ar ions bombardment (especially 72 hours at 350 eV) on poly-crystalline mirror is known to lead to a high roughness causing an increase in the diffuse reflectivity as reported by Voitsenya \textit{et al.} \cite{Voitsenya1}. The high diffuse reflectivity may exclude poly-crystalline Mo mirrors for systems which require mirror cleaning.

\begin{figure}[h]
\centering
\includegraphics[width=0.4\textwidth]{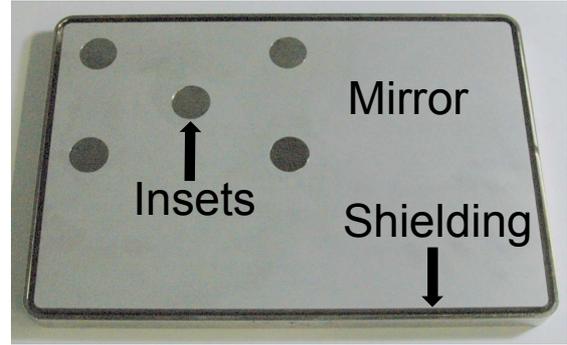}
\caption{\label{TS}Picture of the ITER's edge Thomson Scattering mirror mock-up. 5 Mo mirrors can be inserted.}
\end{figure}

The next step to demonstrate the feasibility of FMs plasma cleaning is to apply it to the mock-up of ITER Edge Thomson Scattering mirror (200$\times$300 mm, Fig \ref{TS}). This mock-up was designed with a shielding to avoid edge effects. The mirror itself is composed of stainless steel with 5 polished NcMo insets to ease the characterization. These experiments were just started.

\subsection{Cleaning in the presence of a magnetic field}
\label{WMF}

\begin{figure}[h]
\centering
\includegraphics[width=0.45\textwidth]{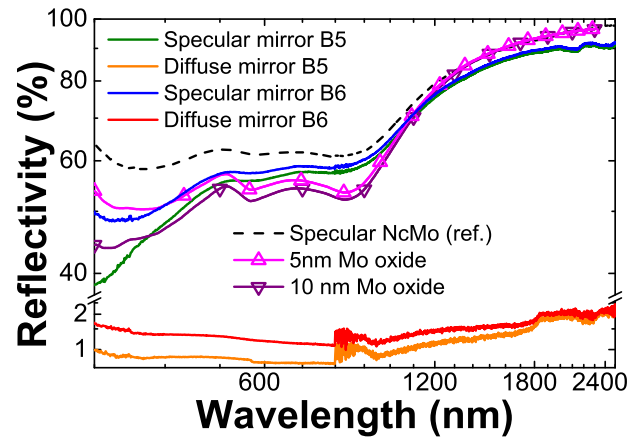}
\caption{\label{SSMo12refl}Specular and diffuse reflectivity of mirror B5 and B6 cleaned for 8h15 and $\alpha$ equal to 90 and 45$^{\circ}$. A reference reflectivity of a nano-crystalline Mo film and reflectivity calculated for a Mo oxide film (5 nm and 10 nm thick) are plotted for comparison.}
\end{figure}

For this experiment 6 mirrors B were coated with a dense Al$_{2}$O$_{3}$ film. 4 were coated with a 5 nm thick film (mirror B1 - B4) and 2 were coated with a 50 nm thick film (mirror B5, B6). Only mirror B1 was characterized by XPS (table \ref{Table}) after the coating and the cleaning was performed without magnetic field. The remaining 5 samples were cleaned in a magnetic field environment (0.35 T) and characterized by XPS. The cleaning of mirror B1 was done in an Ar environment (1.5 Pa) and ion energy of 200 eV as reference. After 2h30 cleaning, 25\% Mo was present on the surface and only 6\% Al was left. This cleaning time served as reference for the other samples. As seen in the table \ref{Table}, the Al$_{2}$O$_{3}$ film was completely removed for mirrors B2 and B3 for $\alpha$ equal to 90$^{\circ}$ and 45$^{\circ}$, respectively. As the samples were in air after cleaning, the Mo surface was oxidized and adsorbed carbon (C) was measured. Fitting of the Mo3d XPS spectrum revealed 2 oxide components: MoO$_{2}$ (229.6 eV) and MoO$_{3}$ (232.4 eV) \cite{Eren3}. For mirror B4, the cleaning was done with the field lines parallel to the mirror's surface (0$^{\circ}$). To achieve a self-bias of 200 V, the RF power was increased from 20 to 145W. The plasma was only stable for 50 minutes and XPS measurements (not shown here) revealed that the surface was pure stainless steel, i.e. the deposited Al$_{2}$O$_{3}$ film and the NcMo coating were removed. For the moment this result is not understood. The cleaning of thicker films was carried out for 8h30 at 90$^{\circ}$ and 45$^{\circ}$ for mirror B5 and B6, respectively. In comparison to previous cleaning, the surface was more oxidized but the Al$_{2}$O$_{3}$ was fully removed. The specular reflectivity of these two mirrors was below the reference as seen in Fig \ref{SSMo12refl}. The calculated reflectivity of a Mo surface with a 5 and 10 nm Mo oxide film on top is also plotted. The similar reflectivity curves and the XPS results confirmed the oxidation after cleaning. The diffuse reflectivity (Fig. \ref{SSMo12refl}) was below 3\% indicating no roughening of the mirror after cleaning. This effect may be due to two reasons: the lower energy of Ar ions and the nano-crystalline structure of mirror B rather than a poly-crystalline mirror (like mirror A).

\begin{table*}
\begin{center}

  \caption{\label{Table}Atomic concentration (in \%) measured by XPS on the surface of the samples before and after cleaning. N.A. stands for non-measured samples.}
  \includegraphics[width=0.75\textwidth]{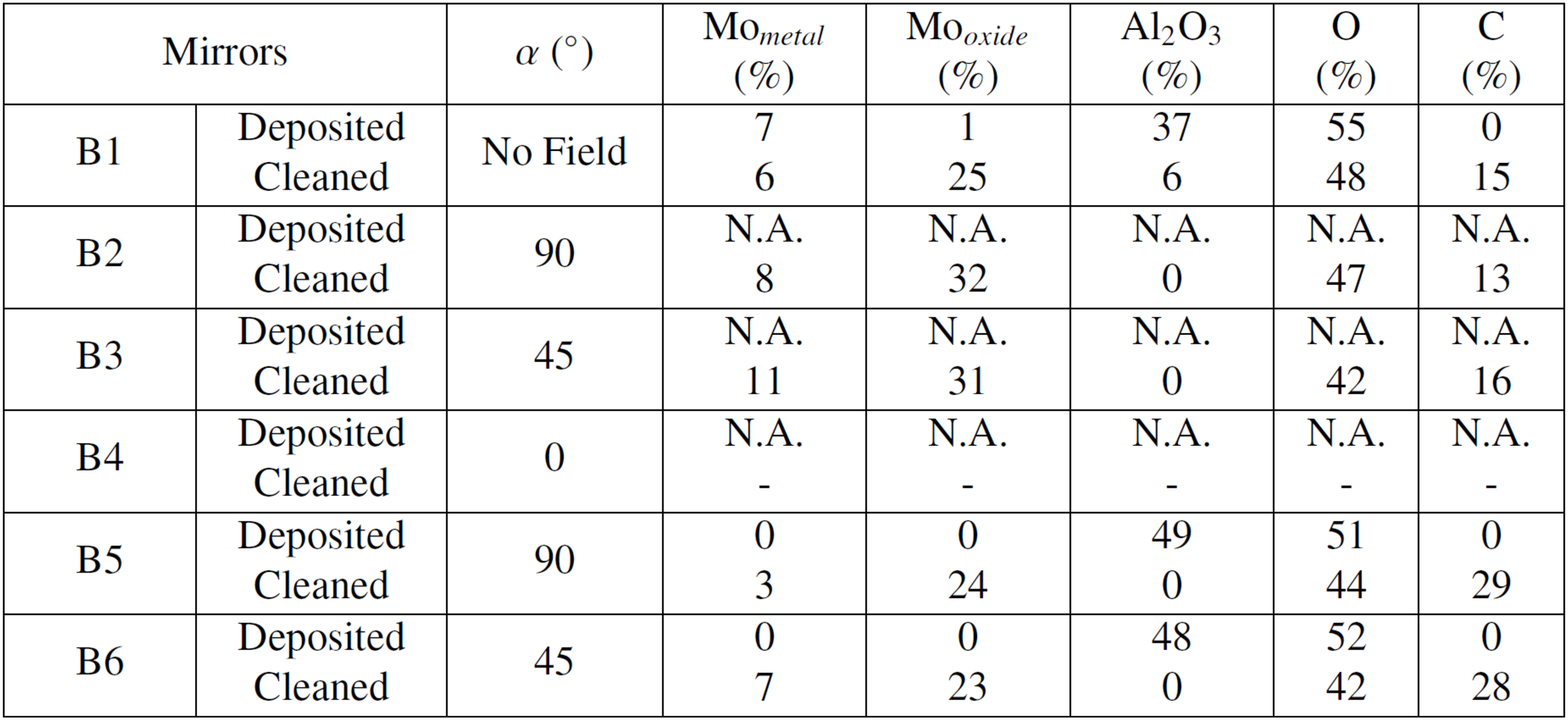}
  \end{center}
\end{table*}

\begin{figure*}[h]
\centering
\includegraphics[width=0.75\textwidth]{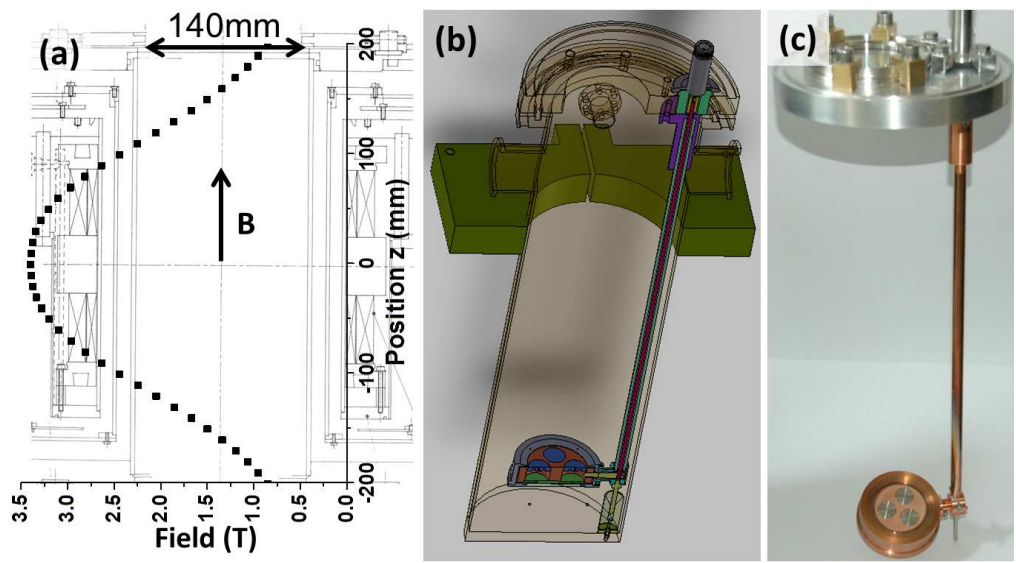}
\caption{\label{Lausanne}(a) Schematic of the superconducting magnet including the magnetic field intensity profile. (b) Cross section of the vacuum chamber and electrode used for the plasma cleaning experiments. (c) Illustration of the electrode (Cu) with 3 mounted mirrors (Mo).}
\end{figure*}

To validate the cleaning procedure in a higher magnetic field, a new vacuum chamber was designed and will be installed in a superconducting magnet used to operate a gyrotron located at the CRPP Lausanne (Fig. \ref{Lausanne}) \cite{Alberti}. Depending on the depth on where the chamber will be inserted in the superconducting magnet, the magnetic field will vary between 1 and 3.5 T. The rotatable electrode provides the possibility to perform experiments for various angles $\alpha$. This project is ongoing and first results will be presented soon.

\section{Conclusion}
\label{4}

A 98 mm diameter mirror with deposits has been cleaned using RF Ar plasma without magnetic field and a bias up to 350 V. Being a poly-crystalline Mo mirror, the diffuse reflectivity increased drastically under the ion bombardment, highlighting the need for single or nano-crystalline materials for ITER's FMs. In a magnetic field (0.35 T), Al$_{2}$O$_{3}$ films were removed from Mo mirrors with Ar RF plasma and for several mirror's orientation to the field. A decrease of the optical performance was observed, mainly due to oxidation of the mirror's surface. The cleaning performance seems to be enhanced when the field lines are parallel (within a few degrees) to the mirror surface. Experiments with a 200$\times$300 mm mock-up mirror and also cleaning in a magnetic field of 3.5 T were started. 

An other important issue is to validate the cleaning process on mirrors deposited with Be and W (laboratory and tokamak deposits): first tests were started in the JET beryllium handling facility under an EFDA Fusion Technology 2013 task and are looking promising for JET-ILW mirrors.

\section*{Acknowledgments}

The authors would like to thank P. Bruzzone and his team for the possibility to work with the Sultan facility at EPFL-CRPP Villigen. This work was supported by the ITER Organization under Contract number 4300000557, 4300000852 and 4300000953. The views and opinions expressed herein do not necessarily reflect those of the ITER Organization. The Swiss Federal Office of Energy, the Federal Office for Education and Science, the Swiss National Foundation (SNF) and the Swiss Nanoscience Institute (SNI) are acknowledged for their financial support.

\label{5}












\end{document}